\documentclass[10pt]{iopart}
\usepackage{iopams}  
\usepackage{graphicx}
\begin{document}

\title{Integrated Plasmonic Metasurfaces for Spectropolarimetry}

\author{Wei Ting Chen$^{1}$, Peter T\"or\"ok$^{2}$, Matthew R. Foreman$^3$, Chun Yen Liao$^{1}$, Wei-Yi Tsai$^{1}$, Pei Ru Wu$^{1}$, 
and Din Ping Tsai$^{1,4}$
}

\address{$^1$Department of Physics, National Taiwan University, Taipei 10617, Taiwan}
\address{$^2$Blackett Laboratory, Department of Physics, Imperial
  College London, London, SW7 2BW, UK}
\address{$^3$Max Planck Institute for the Science of Light,
G\"unther-Scharowsky-Stra{\ss}e 1, 91058 Erlangen, Germany}
\address{$^4$Research Center for Applied Sciences, Academia Sinica, Taipei 11529, Taiwan}
\ead{peter.torok@imperial.ac.uk}
\ead{dptsai@sinica.edu.tw}

\vspace{10pt}
\begin{indented}
\item[]\today
\end{indented}

\begin{abstract}
Plasmonic metasurfaces enable simultaneous control of the phase, momentum, amplitude and polarisation of light and hence promise great utility in realisation of compact photonic devices. In this paper, we demonstrate a novel chip-scale device suitable for simultaneous polarisation and spectral measurements through use of six integrated plasmonic metasurfaces (IPMs), which diffract light with a given polarisation state and spectral component into well-defined spatial domains. Full calibration and characterisation of our device is presented, whereby good spectral resolution and polarisation accuracy over a wavelength range of 500-700~nm is shown. Functionality of our device in a M\"uller matrix modality is demonstrated through determination of the polarisation properties of a commercially available variable waveplate. Our proposed IPM is robust, compact and can be fabricated with a single photolithography step, promising many applications in polarisation imaging, quantum communication and quantitative sensing.
\end{abstract}

%
\vspace{2pc}
\ioptwocol

\section{Introduction}
Optics is an important tool to quantitatively interrogate a broad variety of physical systems and processes, information of which can be extracted through analysis of the optical intensity, polarisation and wavelength of light. Analysis of the polarisation of light can, for example, furnish details regarding sample structure \cite{Camacho2015} and material properties \cite{Hall2013}, whereas spectroscopic measurements carry information regarding chemical composition \cite{Colthup2012}, particle velocity \cite{Leutenegger2011}, local magnetic and electric fields \cite{Donati2009,Bublitz1997} and more \cite{Demtroder2013}. Simultaneous polarisation and spectral measurements, i.e. spectropolarimetry, combines the power of both techniques and constitutes a more exhaustive analytic tool since the information carried by these channels are typically uncorrelated. Accordingly such measurements have seen employ in many branches of science including for example astrophysics \cite{delToroIniesta2003,Sterzik2012}, cosmology \cite{Giovannini2010}, remote sensing \cite{Tyo2006}, biophysics \cite{Volkenstein2012} and data storage \cite{Gu2009}.

Spectropolarimetry can be practically realised using a number of experimental architectures, which differ through the means of spectral and polarisation discrimination. Seperation of individual spectral components is typically achieved  by means of either a tunable optical filter \cite{Gupta2002} or a dispersive element, such as a prism or grating structure \cite{Gorodetski14}. Differing polarisation components, on the other hand, can be found using sequential measurements with variable waveplates and analysers \cite{Pust2006} or simultaneously by means of division of amplitude polarimeters \cite{Azzam1982} or polarization gratings \cite{Gori1999,Kudenov15}. So-called channeled spectropolarimeters represent a further option in which  interferometrically generated carrier frequencies are amplitude modulated according to the state of polarisation \cite{CravenJones2011}, thereby allowing both spectral and polarisation information to be obtained in a single snapshot \cite{Sabatke2002}.

Each spectropolarimeteric architecture brings its own set of advantages and disadvantages according to the specific application at hand (see e.g. \cite{Tyo2006} for a fuller discussion). For example, sequential measurements using variable elements (e.g. filters or polarisation analysers) are unsuitable for scenarios in which the spectral or polarisation content can vary over the course of a measurement. Commonly, such elements are solid-state or liquid crystal based which can be both costly and bulky. Additionally, use of rotating or variable elements can introduce mechanical vibrations, wear and a potentially undesirable energy overhead. The working regime of liquid crystal based elements is also generally limited to the visible or near-infrared \cite{Kudenov15,Zhao22} and thus can not be employed in the infra-red or terahertz regime.  Channeled spectropolarimeters avoid the need for sequential measurements however require more involved data analysis and hence computational resources, and moreover, suffer from a loss of spectral resolution due to use of multiple spectral bands to derive the complete polarisation state \cite{CravenJones2011}. Grating based systems, similarly, allow simultaneous measurements but can suffer from lower signal levels and can involve complicated fabrication processes making them less suitable for device integration.

Plasmonic metasurfaces, i.e. nano-structured thin metallic films, are promising candidates for development of compact photonic devices for spectropolarimetry, since they afford simultaneous control over the phase, momentum, amplitude and polarisation of light \cite{Arbabi1,Meinzer2,Yu2014}. These attributes have accordingly attracted much interest with subsequent realisation of many metasurface-based optical elements, such as beam steerers, polarisers, waveplates, modulators,
holograms and others \cite{Yu4,Pors5,Lin6,Lin7,Kang8,Sun9,Huang10,Chen11,Huang12,Fan13}. Polarisation selective plasmonic metasurfaces have also been reported \cite{Benetou16,Drezet17,Afshinmanesh18,Wen21}, with one shot polarisation measurements  using three interweaved plasmonic metasurfaces reported \cite{Pors23} during the preparation of this article. In this work, we propose and demonstrate the use of six integrated plasmonic metasurfaces (IPMs) for spectropolarimetry hence extending the functionality of earlier designs. Our IPM design, which is detailed in  Section~\ref{sec:principle}, allows many of the limitations of conventional spectropolarimetry methods to be overcome. Moreover, due to its compact size our IPM device allows for direct on-camera integration \cite{Drezet17,Afshinmanesh18,Xie19,Peltzer20}. In Section~\ref{sec:fabrication} we describe the fabrication process of our IPM device which is fully compatible with today's semiconductor manufacturing technologies, before detailing its characterisation over the wavelength range of 500-700~nm in Section~\ref{sec:results}. Finally, we experimentally demonstrate the functionality of the IPM device through determination of the polarisation properties of a commercially available waveplate, before concluding in Section~\ref{sec:conclusions}.

\section{Principle and design \label{sec:principle}}
\begin{figure}[t!]
\begin{center}
\includegraphics{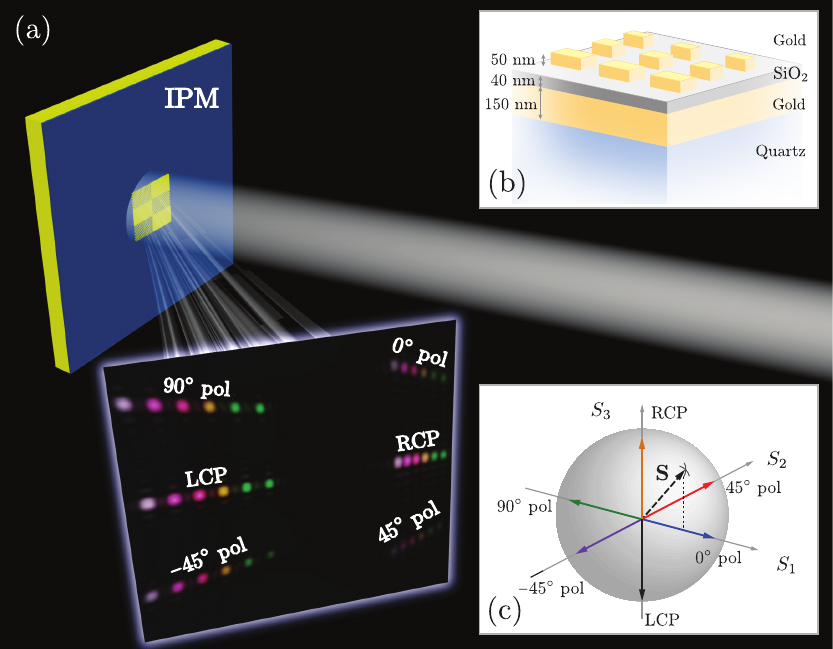}
\caption{Illustration of our IPM device. A $2 \times 3$ array of
  metasurfaces acting as differing polarisation state analysers
  separates input light into its constituent polarisation and spectral
  components. Experimental image shown was obtained for input
  $-45^\circ$ linearly polarised light with 6 spectral components. (b)
  Quasi-planar structure of each metasurface. (c) Schematic of
  polarimetric ``triangulation'' in polarisation space, whereby a given
  polarisation state (dashed arrow) is projected onto a basis of six
  measurement states (solid coloured arrows). Note that  $S_1^2 + S_2^2
  + S_3^2 \leq S_0^2$ such that Stokes vectors must lie within the so-called Poincar\'e sphere.}\label{fig:1}
\end{center}
\end{figure}

\begin{figure}[t]
\begin{center}
\includegraphics{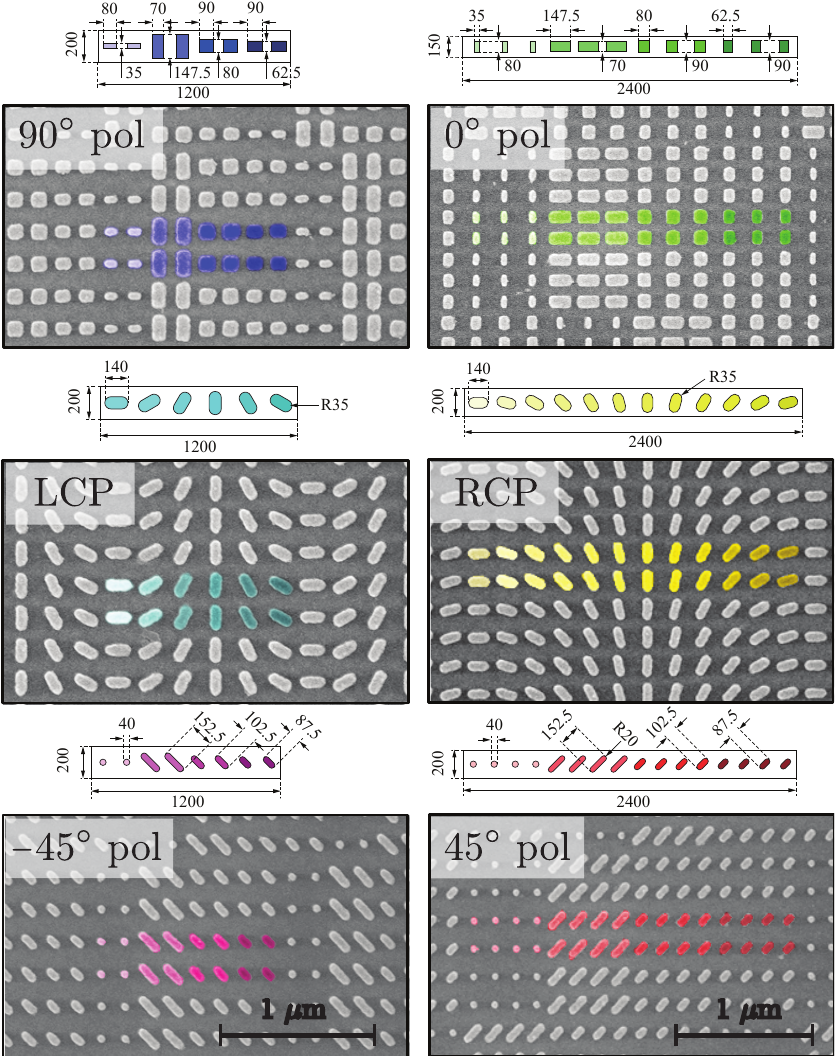}
\caption{Main panels show scanning electron microscope images of regions of the fabricated plasmonic metasurfaces and are labeled with the corresponding polarisation state analysed by each. Schematics of the nominal designs of a single unit cell are also shown. A pseudo-color gradient has been superimposed to depict the linear phase gradient across the unit cell at $\lambda = 700$~nm}
\label{fig:2}
\end{center}
\end{figure}

Complete characterisation of (possibly partially) transversally homogeneously polarised light of a given wavelength can only be achieved via
analysis of the relative intensity of multiple distinct polarisation
components. Practical realisation of a spectropolarimetric metadevice
therefore requires steering of different input polarisation and spectral components into distinct directions as depicted schematically in Figure~\ref{fig:1}(a). Polarisation selective diffractive steering can be achieved using plasmonic metasurfaces consisting of suitably designed nanorod arrays (see Figures~\ref{fig:1}(b)  and \ref{fig:2}) as detailed below, whilst the intrinsic dispersive behavior automatically affords angular multiplexing of different spectral components. Upon collection of the light diffracted
from the IPM, e.g. using a camera or multiple photodiodes, the polarisation
state, parameterised by a Stokes vector $\vec{S} = (S_0,S_1,S_2,S_3)$ can be determined as a function of wavelength \cite{Foreman24} (see Figure~\ref{fig:1}(c)). A minimum of four measurements
are required to ensure that the retrieval of the polarisation
information is not under-determined, therefore requiring at least four metasurfaces. Device performance, however, scales with the number of metasurfaces \cite{Foreman25,Foreman26} due to increased redundancy in the measurements, such that we elect to use an optimal
design consisting of six metasurfaces
arranged in a $2 \times 3$ array corresponding to horizontal (0$^\circ$), vertical
(90$^\circ$), $\pm 45^\circ$, right (RCP) and left circular polarisation (LCP)
analysers (see Figures~\ref{fig:1} and \ref{fig:2}).

Each metasurface consists of rows of gold nanorods arranged periodically in the horizontal ($x$) direction (with period $\Lambda$) as shown in Figure~\ref{fig:2} and patterned on a 40~nm thick SiO$_2$
layer on a gold mirror, yielding a total thickness of 260~nm excluding
substrate (see Figure~\ref{fig:1}(b)). A reflection modality was adopted due to the
greater optical efficiency it confers and the thickness of the SiO$_2$
chosen so as to maximise the reflectance of the linear polarisation
channels, which is lower than in the circular channels \cite{Pors5,Pors27,Zheng28,Ding29}. We note that although metasurfaces approximately 72~$\mu\mbox{m} \times 48$~$\mu$m in size were used, the dimensions are scalable to 2.4~$\mu\mbox{m} \times 2.4$~$\mu$m by reduction of the number of constituent unit cells to help avoid potential spatial coherence related issues. 


Interaction of an incident electromagnetic wave with each metasurface can be modeled using a generalised Snell's law \cite{Yu2014}, whereby, $k n_i (\sin \theta_r - \sin\theta_i) = d\Phi/dx$ where $k = 2\pi/\lambda$ is the vacuum wavenumber, $\lambda$ is the wavelength, $n_i$ is the refractive index of the surrounding medium, $\theta_{i,r}$ are the angles of incidence and reflection respectively and $\Phi$ is the interfacial phase discontinuity imposed by the metasurface upon reflection. Continuous tuning of the phase discontinuity can be achieved across each unit cell by variation of the length, width or orientation of each constituent nanorod, which in turn controls the spectral position and excitation strength of the plasmonic eigenmodes (see Supporting Information 1). To steer the incident light into a well-defined direction, we therefore engineer the phase anisotropy so that a specific incident polarisation sees a constant gradient  $d\Phi/dx = 2\pi/\Lambda$. Full three dimensional finite element simulations were used to identify suitable structural parameters for each nanorod (see Supporting Information 1 and 2). The final designs are shown in Figure~\ref{fig:2}. For metasurfaces acting as linear analysers, only the length and width of each nanorod was varied (four distinct geometries were used in each unit cell), whilst for the LCP and RCP channels $\Phi$ was controlled through variation of the nanorod orientation (6 and 12 different orientations uniformly distributed over 180$^\circ$ were used respectively). Although the linear phase gradient was designed for a nominal wavelength, it exhibits only a weak wavelength dependence, allowing broadband operation as has been previously demonstrated \cite{Yu4,Sun9,Pors23}. As a consequence of the linear phase gradient, normally incident light of the desired polarisation is diffracted at an angle $\theta_r = \sin^{-1}(\lambda/\Lambda)$, where we have assumed $n_i = 1$ to match our experimental conditions. Calculations of the associated extinction ratios can be found in Supporting Information 3. Importantly, a dispersive behavior is exhibited such that angular multiplexing of different spectral components is automatically achieved. Angular discrimination of the different polarisation components originating from each metasurface was achieved by choosing differing periods of $\Lambda = 1.2$ and 2.4~$\mu$m for metasurfaces on the left and right side of the metasurface array. Similarly, a vertical periodicity of 4.8~$\mu$m was imposed for the top and bottoms rows to achieve a vertical displacement in the output beams.

\section{Fabrication and characterisation\label{sec:fabrication}}

\begin{figure*}[t]
\begin{center}
\includegraphics{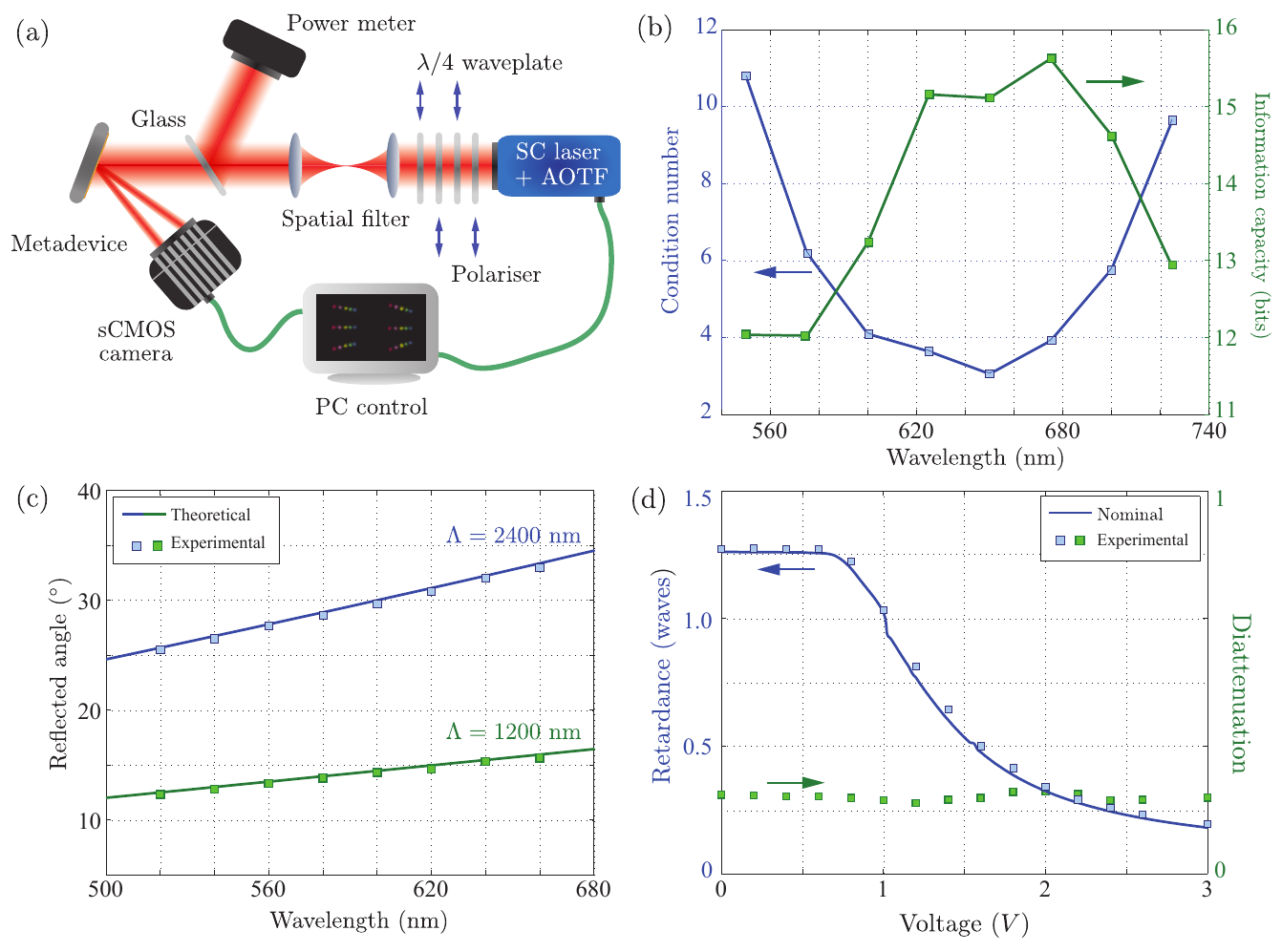}
\caption{(a) Experimental setup used for characterisation of our IPM device. SC: supercontinuum laser. AOTF: acousto-optic filter (b) Variation of condition number and information capacity as quantified by the logarithm of the number of distinguishable polarisation states. (c) Theoretical predictions and experimentally observed spectral dispersion for circular detection channels. (d) Retardance and diattenuation values extracted from measured Müller matrices of a known variable waveplate.}
\label{fig:3}
\end{center}
\end{figure*}
Fabrication of our IPM metadevice was performed using a standard e-beam lithography and lift-off process. Specifically a 3~nm thick gold (Au) film was first sputtered onto a quartz wafer. This was followed by thermal evaporation of another 147~nm Au layer to bring the total thickness of the gold layer to 150~nm. The sample was subsequently spin coated with a 260~nm thick SiO$_2$ (IC1-200, from Futurrex, Inc) layer and baked for 3 min on a 200$^\circ$C hotplate. Subsequently, we used reactive ion etching with tetrafluoromethane (CF$_4$) to reduce the thickness to 40~nm. The sample was then further coated with an e-beam resist (ZEP 520A from ZEON Corporation) of 180~nm thickness, and baked for 3~min on a hotplate at 180$^\circ$C. Thereafter the individual plasmonic metasurfaces were defined using an e-beam lithography system (Elionix ELS-7000) using an acceleration voltage of 100~keV and current of 30~pA. After exposure, the sample was developed in a solution of ZED N50 (ZEON Corporation) for 180~s. Once the development of the resist was complete, a 50~nm thick Au film was deposited by electron-beam evaporation. The sample was then soaked in ZEMAC (ZEON Corporation) for over 12~hrs, before the unpatterned regions were finally removed in an ultrasonic cleaner.

Following fabrication we characterised the metadevice using the experimental setup shown in Figure~\ref{fig:3}(a). Specifically, we used a supercontinuum laser combined with an acousto-optic filter, which outputs a series of discrete, equally spaced laser peaks, i.e. a frequency comb. The desired polarisation state was generated by introduction of several polarisers and phase retarders. A thin, uncoated glass plate was also inserted into the collimated light path at near normal incidence to provide reference intensity measurements for device calibration. Light diffracted by the IPM was directly collected by a sCMOS camera. Figure~\ref{fig:1} shows a typical experimental image when the IPM device is illuminated with $-45^\circ$ polarised light. As the incident polarisation is varied, so the relative intensity of each intensity peak changes, as can be seen in the Supplementary Movie. To study the polarisation response of our metadevice for a given wavelength, we must determine the relationship between the intensity of individual peaks arising from diffraction from each metasurface (which we collect into a vector of intensities $\vec{I}$) and the incident state of polarisation $\vec{S}$, which can be written in the form $\vec{I} = \mathbf{A}\vec{S}$  where $\mathbf{A}$ is a $6 \times 4$ instrument matrix. To determine $\mathbf{A}$ we followed the procedure detailed in \cite{Azzam30} and Supporting Information 4. This procedure also acts as a calibration for subsequent spectropolarimetric measurements and must be performed for all wavelengths. As an example the experimental instrument matrix found at $\lambda = 650$~nm is 
\begin{equation*}
\mathbf{A} = \left(\begin{array}{cccc} 
0.2754&0.2287&0.0069&-0.0078\\
0.2447&-0.2033&0.0001&-0.0063\\
0.4750&0.0037&0.0582&-0.4584\\
0.4484&0.0183&-0.0416&0.3928\\
0.2210&-0.0005&0.1709&0.0059\\
0.2747&0.0110&-0.2224&0.0245
\end{array}\right) 
\end{equation*}
from which it is apparent that the first two rows correspond to a 0$^\circ$ and 90$^\circ$ linear analyser respectively, the third and fourth to LCP and RCP analysers and finally the fifth and sixth row to $+45^\circ$ and $-45^\circ$ analysers, therefore matching the nominal IPM design. Approximately double the power is found in the RCP and LCP channels as compared to the linear channels. This is predicted from numerical calculations of the diffraction efficiencies which are $\sim 36$\% and 55\% for the linear and circular channels respectively. Experimental instrument matrices for alternative wavelengths are given in Supporting Information 5.

\section{Results and discussion\label{sec:results}}
Critically, the instrument matrix affects a number of key performance
metrics for the IPM device. Specifically, the condition number of the
instrument matrix, $\kappa = \|\mathbf{A}\| \| \mathbf{A}^{-1}\|$ where $\|\cdot \|$ denotes the $p = 2$ matrix norm, quantifies
the maximum extent to which noise is amplified during the polarisation
reconstruction process. Smaller condition numbers are preferable, with
our metadevice exhibiting a minimum of 3.028 at $\lambda = 650$~nm (Figure~\ref{fig:3}(b)). Theoretically the minimum achievable condition number is $\kappa = \sqrt{3} \approx 1.732$,
corresponding to the nominal design of our metadevice, albeit with
balanced detectors \cite{Foreman26}. When the imbalance arising from differing
diffraction efficiencies is considered the theoretical condition
number of our IPM is $\approx 2.082$. 

The instrument matrix, in concert with experimental (noisy) intensity
data, can also be used to quantify the polarisation resolution and
hence information capacity of our metadevice. The former describes the
precision to which a polarisation state can be determined, whilst the
latter dictates the average information gain per measurement (for each
wavelength channel and a given data acquisition rate). Calculation of
the polarisation resolution followed an information theoretic approach
based on the Cramer-Rao lower bound \cite{Foreman24}. Using this method the volume
of uncertainty in polarisation space was first calculated from the
associated Fisher information matrix, from which the total number of
distinguishable polarisation states was derived. A Shannon information
capacity then follows (see Supporting Information 6 for further
details) and is shown in Figure~\ref{fig:3}(b). At wavelengths $\sim 650$~nm, the
information per measurement of $\sim 15$~bits is largest, however drops to
$\sim 12$~bits per measurement away from this optimal wavelength.

Of further importance for all spectroscopic measurements is the spectral resolution. For our device this is closely related to the angular dispersion, $\alpha$, and hence period, $\Lambda$, of each metasurface. Figure~\ref{fig:3}(c) compares the theoretical and experimentally measured angular dispersions for the LCP and RCP channels ($\Lambda = 1200$ and 2400~nm respectively). Excellent agreement is evident in both cases with $\alpha \approx 0.053$ and 0.024$^\circ$/nm. Assuming we use our device as the dispersive element in a spectrometer with a focal length of 25~cm focal length and use a CCD detector with $N = 2048$~pixels (pixel width $W_p = 14$~$\mu$m and total width of 28.67~mm), the wavelength spans, $\Delta\lambda$, on the CCD are 123.85 and 273.8~nm respectively. Finally, the spectral resolution follows as $\delta\lambda = (RF \cdot \Delta \lambda \cdot W_s)/(N \cdot W_p)$, where $W_s$ is the entrance slit width and the resolution factor $RF$ is determined by the relationship between slit width and the pixel width \cite{Vollmann31}. Assuming typical values of $W_s \approx 2W_p$ and $RF \approx 2.5$ we find spectral resolutions of 0.3 and 0.67~nm for the LCP and RCP channels. Similar spectral resolutions for the (90$^\circ$, $-45^\circ$) and (0$^\circ$, $+45^\circ$) channels are also found.

Once the system is characterised, spectropolarimetric measurements can readily be performed. The IPM can either be used in the calibration setup or built into a new instrument, such as an ellipsometer or a polarised light microscope. Introduction of a sample before the IPM, however, can modify the incident polarisation state as described by the M\"uller matrix, $\mathbf{M}(\lambda)$, of the sample. With knowledge of the incident polarisation the change of polarisation can be inferred since the measured intensities are given by $\vec{I}(\lambda) = \mathbf{A}(\lambda) \mathbf{M}(\lambda)\vec{S}$. Moreover, by illuminating the sample-IPM combination with four (or more) differently polarised beams (parameterised by the Stokes vectors $\vec{S}_j$, $j = 1,\ldots, 4$, which we combine into a matrix $\mathbf{S} = [\vec{S}_1,\vec{S}_2,\vec{S}_3,\vec{S}_4]$), multiple output intensity vectors $\vec{I}_j(\lambda)$  (which we also combine into a matrix $\mathbf{I}(\lambda)$) result. The resulting matrix equation
\begin{equation}
\mathbf{I}(\lambda) = \mathbf{A}(\lambda)\mathbf{M}(\lambda)\mathbf{S}\label{eq1}
\end{equation}
can be simply inverted, therefore allowing determination of the complete polarisation properties of a sample, as contained in the M\"uller matrix. A well-conditioned instrument matrix is essential for the stability of the inversion. Despite the good conditioning of our IPM device, inaccuracies can arise e.g. from errors in incident polarisation states, which in turn give rise to non-physical M\"uller matrices. Physically admissible M\"uller matrices were ensured in our work, however, by adopting a constrained maximum-likelihood estimator \cite{Aiello32} seeded with a least norm solution of Equation~(\ref{eq1}).

To demonstrate the capabilities of our metadevice we have obtained the M\"uller matrix of a liquid crystal tunable full waveplate (Thorlabs LCC1223-A) (see Supporting Information 7). Nominal retardance values at $\lambda = 635$~nm as provided by Thorlabs Inc. are shown in Figure~\ref{fig:3}(d) as a function of applied voltage. Extraction of the retardance of the waveplate from experimental M\"uller matrices (acquired at $\lambda = 650$~nm) was achieved through a polar decomposition of the associated M\"uller matrix \cite{Lu33}, which represents an arbitrary element as a series of a pure diattenuator, retarder and depolariser. Experimentally determined retardance values are also shown in Figure~\ref{fig:3}(d) from which good agreement is seen albeit for a small systematic offset attributable to wavelength dependent retardance of the waveplate. A slight parasitic diattenuation is also seen which we believe arises from the liquid crystal nature of the waveplate.

\section{Conclusions\label{sec:conclusions}}
In conclusion, by integration of six plasmonic metasurfaces, each
engineered to respond differently to specific incident polarisation
states, we have realised a compact spectropolarimetric
device. Dispersive properties of our metadevice allowed simultaneous
spectral measurements of incident light to be made, with spectral
resolutions of $\sim 0.3$~nm easily achievable upon insertion of our IPM
device into typical spectrometer setups. Device functionality over a
large bandwidth was verified through experimental characterisation in
terms of the condition number of the associated instrument matrix and
the information capacity. Finally, when used in a M\"uller matrix
polarimetric modality, our IPM device accurately reproduced the known
polarisation properties of a variable waveplate inserted before the
IPM. Due to the integrability, compact size, low energy overhead and
compatibility with current manufacturing techniques, we envisage that
the proposed IPM device can find successful applications in any field
employing the polarisation of light such as quantitative bioscience,
remote sensing, quantum communication and polarisation imaging.

\section*{Acknowledgements}
DPT acknowledges support from the Ministry of Science and Technology,
Taiwan (Grants 103-2745-M-002-004-ASP and 103-2911-I-002-594) and
Academia Sinica (Grant AS-103- TP-A06). PT was funded via an Academia
Sinica Professorial Scholarship. MRF acknowledges support from an
Alexander von Humboldt Fellowship and the Max Planck Society. The
authors also thank Dr. C. Macias-Romero (\'Ecole Polytechnique F\'ed\'erale
de Lausanne, Switzerland) for discussions.


\section*{References}
\begin{thebibliography}{99}

\bibitem{Camacho2015} Camacho R, Tubasum S, Southall J, Cogdell R J, Sforazzini G, Anderson H L, Pullerits T and Scheblykin I G 2015 Fluorescence polarization measures energy funneling in single light-harvesting antennas-LH2 vs conjugated polymers,
Scientific Reports \textbf{5},15080 

\bibitem{Hall2013} Hall S A, Hoyle M-A, Post J S, and Hore D K 2013 Combined Stokes Vector and Mueller Matrix Polarimetry for Materials Characterization, Anal. Chem. \textbf{85} 7613--7619

\bibitem{Colthup2012} Colthup, N 2012 \emph{Introduction to infrared and Raman spectroscopy} Elsevier

\bibitem{Leutenegger2011} Leutenegger M, Martin-Williams E, Harbi P, Thacher T, Raffoul W, Andr\'e M, Lopez A, Lasser P, and Lasser T 2011 Real-time full field laser Doppler imaging, Biomed. Opt. Express \textbf{2} 1470-1477 

\bibitem{Donati2009} Donati J-F and Landstreet J D 2009 Magnetic Fields of Nondegenerate Stars, Ann. Rev. Astron. Astro. \textbf{47} 333-370 

\bibitem{Bublitz1997} Bublitz, G U, and Boxer S G 1997 Stark spectroscopy: applications in chemistry, biology, and materials science, Ann. Rev. Phys. Chem. \textbf{48} 213-242

\bibitem{Demtroder2013} Demtr\"oder, W 2013 \emph{Laser spectroscopy: basic concepts and instrumentation} Springer Science \& Business Media, Berlin

\bibitem{delToroIniesta2003} del Toro Iniesta, J C 2003 \emph{Introduction to spectropolarimetry} Cambridge University Press


\bibitem{Sterzik2012} Sterzik M F, Bagnulo S, and Palle E, Biosignatures as revealed by spectropolarimetry of Earthshine 2012 Nature \textbf{483}, 64-66

\bibitem{Giovannini2010} Giovannini, M 2010 Circular dichroism, magnetic knots, and the spectropolarimetry of the cosmic microwave background, Phys. Rev. D \textbf{81} 023003

\bibitem{Tyo2006} Tyo J S, Goldstein D L, Chenault D B, and Shaw J A, 2006 Review of passive imaging polarimetry for remote sensing applications, Appl. Opt. \textbf{45} 5453-5469

\bibitem{Volkenstein2012} Volkenstein, M V 1977 \emph{Molecular Biophysics} Academic Press, New York

\bibitem{Gu2009} Chon J, Zijlstra P, and Gu M 2009 Five-dimensional optical recording mediated by surface plasmons in gold nanorods, Nature \textbf{459}, 410-413




\bibitem{Gupta2002} Gupta N, Dahmani R, and Choy S 2002 Acousto-optic tunable filter based visible- to near-infrared spectropolarimetric imager, Opt. Eng. \textbf{41} 1033-1038

\bibitem{Gorodetski14} Gorodetski Y, Biener G, Niv A, Kleiner V and Hasman E 2005 Space-variant polarization manipulation for far-field polarimetry by use of subwavelength dielectric gratings, Opt. Lett. \textbf{30} 2245-7


\bibitem{Pust2006} Pust N J  and Shaw J A 2006 Dual-field imaging polarimeter using liquid crystal variable retarders, Appl. Opt. \textbf{45} 5470-5478

\bibitem{Azzam1982} Azzam R M A 1982 Division-of-amplitude photopolarimeter (DOAP) for the simultaneous measurement of all four Stokes parameters of light, Opt. Acta \textbf{29} 685-689


\bibitem{Gori1999} Gori F 1999 Measuring Stokes parameters by means of a polarization grating Opt. Lett. \textbf{24} 584-586


\bibitem{Kudenov15} Kudenov M W, Escuti M J, Dereniak E L and Oka K 2011 White-light channeled imaging polarimeter using broadband polarization gratings, Appl. Opt. \textbf{50} 2283-93

\bibitem{CravenJones2011} Craven-Jones J, Kudenov M W, Stapelbroek M G, and Dereniak E L 2011 Infrared hyperspectral imaging polarimeter using birefringent prisms, Appl. Opt. \textbf{50}, 1170-1185

\bibitem{Sabatke2002} Sabatke D S, Locke A M, Dereniak E L, Descour M R, Garcia J P,  Hamilton T K, and McMillan R W, 2002 Snapshot imaging spectropolarimeter Opt. Eng. \textbf{41} 1048-1054

\bibitem{Zhao22} Zhao X, Bermak A, Boussaid F and Chigrinov V G 2010 Liquid-crystal micropolarimeter array for full Stokes polarization imaging in visible spectrum, Opt. Express \textbf{18} 17776-87

\bibitem{Arbabi1} Arbabi A, Horie Y, Bagheri M and Faraon A 2015 Dielectric metasurfaces for complete control of phase and polarization with subwavelength spatial resolution and high transmission, Nature Nanotech. \textbf{10} 937-43

\bibitem{Meinzer2} Meinzer N, Barnes W L and Hooper I R 2014 Plasmonic meta-atoms and metasurfaces, Nature Photon. \textbf{8} 889-98

\bibitem{Yu2014} Yu N and Capasso F 2014 Flat optics with designer metasurfaces, Nature Mater. \textbf{13} 139-50

\bibitem{Yu4} Yu N, Aieta F, Genevet P, Kats M A, Gaburro Z and Capasso F 2012 A Broadband, Background-Free Quarter-Wave Plate Based on Plasmonic Metasurfaces, Nano Lett. \textbf{12} 6328-33

\bibitem{Pors5} Pors A and Bozhevolnyi S I 2013 Plasmonic metasurfaces for efficient phase control in reflection, Opt. Express. \textbf{21} 27438-51

\bibitem{Lin6} Lin J, Mueller J P B, Wang Q, Yuan G, Antoniou N, Yuan X-C and Capasso F 2013 Polarization-Controlled Tunable Directional Coupling of Surface Plasmon Polaritons, Science \textbf{340} 331-4

\bibitem{Lin7} Lin D, Fan P, Hasman E and Brongersma M L 2014 Dielectric gradient metasurface optical elements, Science \textbf{345} 298-302

\bibitem{Kang8} Kang M, Feng T, Wang H-T and Li J 2012 Wave front engineering from an array of thin aperture antennas, Opt. Express \textbf{20} 15882-90

\bibitem{Sun9} Sun S et al. 2012 High-Efficiency Broadband Anomalous Reflection by Gradient Meta-Surfaces, Nano Lett. \textbf{12} 6223-9

\bibitem{Huang10} Huang L, Chen X, Mühlenbernd H, Li G, Bai B, Tan Q, Jin G, Zentgraf T and Zhang S 2012 Dispersionless Phase Discontinuities for Controlling Light Propagation, Nano Lett. \textbf{12} 5750-5

\bibitem{Chen11} Chen W T et al. 2014 High-Efficiency Broadband Meta-Hologram with Polarization-Controlled Dual Images, Nano Lett. \textbf{14} 225-30

\bibitem{Huang12} Huang L, Chen X, Bai B, Tan Q, Jin G, Zentgraf T and Zhang S 2013 Helicity dependent directional surface plasmon polariton excitation using a metasurface with interfacial phase discontinuity, Light Sci Appl \textbf{2} e70

\bibitem{Fan13} Fan R-H, Zhou Y, Ren X-P, Peng R-W, Jiang S-C, Xu D-H, Xiong X, Huang X-R and Wang M 2015 Freely Tunable Broadband Polarization Rotator for Terahertz Waves, Adv. Mater. \textbf{27} 1201-6


\bibitem{Benetou16} Benetou M I, Thomsen B C, Bayvel P, Dickson W and Zayats A V 2011 Four-level polarization discriminator based on a surface plasmon polaritonic crystal, Appl. Phys. Lett. \textbf{98} 111109

\bibitem{Drezet17} Drezet A, Genet C and Ebbesen T W 2008 Miniature Plasmonic Wave Plates, Phys. Rev. Lett. \textbf{101} 043902

\bibitem{Afshinmanesh18} Afshinmanesh F, White Justin S, Cai W and Brongersma Mark L. 2012 Measurement of the polarization state of light using an integrated plasmonic polarimeter, Nanophotonics \textbf{1}   125.

\bibitem{Wen21} Wen D et al. 2015 Metasurface for characterization of the polarization state of light, Opt. Express \textbf{23} 10272-81

\bibitem{Pors23} Pors A, Nielsen M G and Bozhevolnyi S I 2015 Plasmonic metagratings for simultaneous determination of Stokes parameters, Optica \textbf{2} 716-23

\bibitem{Xie19} Xie Y-B, Liu Z-Y, Wang Q-J, Zhu Y-Y and Zhang X-J 2014 Miniature polarization analyzer based on surface plasmon polaritons, Appl. Phys. Lett. \textbf{105} 101107

\bibitem{Peltzer20} Peltzer J J, Bachman K A, Rose J W, Flammer P D, Furtak T E, Collins R T and Hollingsworth R E. 2012 Plasmonic micropolarizers for full Stokes vector imaging, Proc. SPIE. 83640O-O-12.


\bibitem{Foreman24} Foreman M R and T\"{o}r\"{o}k P 2010 Information and resolution in electromagnetic optical systems, Phys. Rev. A \textbf{82} 043835

\bibitem{Foreman25} Foreman M R, Macias-Romero C and T\"{o}r\"{o}k P 2008 A priori information and optimisation in polarimetry, Opt. Express \textbf{16} 15212-27

\bibitem{Foreman26} Foreman M R, Favaro A and Aiello A 2015 Optimal Frames for Polarisation State Reconstruction, Phys. Rev. Lett. 115, 263901 

\bibitem{Pors27} Pors A, Nielsen M G and Bozhevolnyi S I 2013 Broadband plasmonic half-wave plates in reflection, Opt. Lett. \textbf{38} 513-5

\bibitem{Zheng28} Zheng G, Mühlenbernd H, Kenney M, Li G, Zentgraf T and Zhang S 2015 Metasurface holograms reaching 80\% efficiency, Nature Nanotech. \textbf{10} 308-12

\bibitem{Ding29} Ding F, Wang Z, He S, Shalaev V M and Kildishev A V 2015 Broadband High-Efficiency Half-Wave Plate: A Supercell-Based Plasmonic Metasurface Approach, ACS Nano \textbf{9} 4111-9

\bibitem{Azzam30} Azzam R M A and Lopez A G 1989 Accurate calibration of the four-detector photopolarimeter with imperfect polarizing optical elements, J. Opt. Soc. Am. A \textbf{6} 1513-21

\bibitem{Vollmann31} Vollmann T E A K 2015 \emph{Spectroscopic Instrumentation - Fundamentals and Guidelines for Astronomers} Springer-Verlag

\bibitem{Aiello32} Aiello A, Puentes G, Voigt D and Woerdman J P 2006 Maximum-likelihood estimation of Mueller matrices, Opt. Lett. \textbf{31} 817-9

\bibitem{Lu33} Lu S -Y and Chipman R A 1996 Interpretation of Mueller matrices based on polar decomposition, J. Opt. Soc. Am. A \textbf{13} 1106-13

\end{thebibliography}
\end{document}